\documentclass[prb,aps,showpacs,showkeys]{revtex4-1}
\pdfoutput=1
\usepackage{color}
\usepackage[pdftex]{graphicx,hyperref}
\usepackage{dcolumn}
\usepackage{bm}
\usepackage{marvosym}
\usepackage{times,amsmath,amssymb}

\begin{document}
\title
{Surface Plasmon Resonance in a metallic nanoparticle embedded in a semiconductor matrix: exciton-plasmon coupling}

\author{Rui M. S. Pereira}
\altaffiliation{Centro de F\'{\i}sica \& Departamento de Matem\'atica, Universidade do Minho, Campus de Gualtar, Braga
4710-057, Portugal}

\author{Joel Borges}
\altaffiliation{Centro de F\'{\i}sica, Universidade do Minho, Campus de Gualtar, Braga
4710-057, Portugal}

\author{Gueorgui V. Smirnov }
\altaffiliation{Centro de F\'{\i}sica \& Departamento de Matem\'atica, Universidade do Minho, Campus de Gualtar, Braga
4710-057, Portugal}

\author{Filipe Vaz}
\altaffiliation{Centro de F\'{\i}sica \& Departamento de F\'{\i}sica, Universidade do Minho, Campus de Gualtar, Braga
4710-057, Portugal}

\author{Mikhail I. Vasilevskiy}
\affiliation{Centro de F\'{\i}sica \& Departamento de F\'{\i}sica, Universidade do Minho, Campus de Gualtar, Braga
4710-057, Portugal}

\affiliation{International Iberian Nanotechnology Laboratory, Braga 4715-330, Portugal}


\begin{abstract}

We consider the effect of electromagnetic coupling between localized surface plasmons in a metallic nanoparticle (NP) and excitons or weakly interacting electron-hole pairs in a semiconductor matrix where the NP is embedded. An expression is derived for the NP polarizability renormalized by this coupling and two possible situations are analyzed, both compatable with the conditions for Fano-type resonances: (i) a narrow bound exciton transition overlapping with the NP surface plasmon resonance (SPR), and (ii) SPR overlapping with a parabolic absorption band due to electron-hole transitions in the semiconductor. The absorption band lineshape is strongly non-Lorentzian in both cases and similar to the typical Fano spectrum in the case (i). However, it looks differently in the situation (ii) that takes place for gold NPs embedded in a CuO film and the use of the renormalized polarizability derived in this work permits to obtain a very good fit to the experimentally measured LSPR lineshape.   

\end{abstract}
\maketitle


\section{Introduction}

Plasmonic nanoparticles (NPs) have received considerable attention of researchers owing to their high sensitivity to the dielectric environment, which is based on the Localized Surface Plasmon Resonance (LSPR) phenomenon \cite{Bohren-Huffman,Stockman2011}. The characteristic LSPR band in the light absorption spectra can be tailored by controlling the size, shape and spatial distributions of the NPs and by choosing the host dielectric matrix where the NPs are embedded or even the substrate on which they are deposited.\cite{Bohren-Huffman,GarciadeAbajo2007,Lachaine2016,Amorim2017,Abid2017} 
It opens a wide range of possibilities for designing novel nanomaterials with the targeted applications including detection of bio-\cite{Anker2008,Cappi2015} and gas\cite{Ghod2011} molecules, plasmonic nano-antennas,\cite{Novotny2011,Schreiber2014} surface-enhanced spectroscopy,\cite{Chumanov,Basko2000,Carnegie2017} LSPR-enhanced solar cells\cite{Quiao2011} as well as systems for nanometrology,\cite{Hill2014} nanolithography and photocatalysis.\cite{Blackman}

A century ago, Gustav Mie deduced a solution of Maxwell’s equations for a sphere, which permit to calculate the scattering and extinction spectra of a metallic nanoparticle,\cite{Bohren-Huffman} The extinction cross-section depends of the dielectric functions of the metal ($\epsilon_{M}$) and host matrix ($\epsilon_h$), and the sphere's radius ($a$) and the wavelength of the incident electromagnetic (EM) wave ($\lambda$). In the limit $a/\lambda \ll 1$ (relevant to NPs), Mie's expressions   are greatly simplified because the sphere's response is dominated by its dipole  contribution (compared to higher multipoles). In this limit, the particle's polarizability becomes independent of its size (unless $\epsilon_{M}$ is size dependent because of quantum confinement or other effects) and given by\cite{Bohren-Huffman}
\begin{equation}
\alpha_0 (\omega)=\frac{\epsilon_M (\omega) - \epsilon_{h}}{\epsilon_M (\omega) +2 \epsilon_{h}}a^3\, ,
\label{bare_eq1}
\end{equation}
where $\omega $ is the EM field frequency. The dipolar resonance (i.e. LSPR) frequency is determined by the pole of the expression (\ref{bare_eq1}). Obviously, it depends on the host dielectric constant and it is the basic principle of the NPs' use for environmental sensing. Of course, NP's shape has a major impact on the polarizability and resonance frequencies,\cite{Stockman2011} in particular, the former becomes a tensor and the three-fold degeneracy of the single dipolar resonance in a sphere is lifted for less symmetric shapes. However, for the sake of simplicity, here we shall confine our consideration by spherical NPs.

Within the dipole approximation, further effects can be taken into account, such as the EM interactions between the particles, which are relevant since in practice one almost always deals with an ensemble of them. These interactions were considered for the first time by J. C. Maxwell-Garnett\cite{MG} and are one of the most important effects influencing the LSPR lineshape, sensitive to the spatial distribution of the particles (such as short-range clustering or formation of larger aggregates, eventually of fractal dimension).\cite{Shalaev,Pereira2013,Vieaud2018} Not only the spectral position but also the width and the shape of the LSPR band is important for applications.  For instance, it has been noticed that the response of LSPR band curvature to $\epsilon _h$ changes is superior to peak shifting in terms of signal-to-noise ratio.\cite{Chen2014}
A broad SPR band facilitates the requirement to couple with the excitation wavelength used in Surface Enhanced Raman Scattering (SERS) detectors and can provide a double resonance at both excitation and Stokes-shifted frequencies\cite{Chu2010} but, on the other hand, steep resonances are better for detection of spectral shifts in the refractive index change--based molecular sensing. Both situations can be achieved with plasmonic NPs, moreover, strongly asymmetric bands due to Fano resonances\cite{Fano} can be achieved in finite clusters (oligomers) of plasmonic particles for sufficiently small interparticle separations.\cite{Lukyanchuk2010,Note2} 

Interaction with other excitations in surrounding bodies that may be in resonance with the localized surface plasmons should also affects the characteristics of the LSPR band. Some oxide semiconductors are potentially interesting materials for using as a matrix for embedding gold NPs. For instance, CuO films with embedded nanoparticles have been demonstrated to be good for CO sensing.\cite{Ghod2011,Proenca2017}  CuO is a semiconductor with a band gap of $\approx 1.5$ eV.\cite{CuO} That is, such a matrix absorbs light itself, via excitonic and inter-band transitions 
and there are dynamical dipoles associated with these transitions, which must couple to the localized surface plasmons.
How will it affect the LSPR band position and line shape due to embedded NPs?
Can we simply replace the constant $\epsilon _h$ by an appropriate complex function $\epsilon _h (\omega)$ in the the expression (\ref{bare_eq1}) for NP’s polarizability or the effect of an absorbing matrix requires a more sophisticated description? 

In this paper we show that the NP's polarizability is renormalized because of the electromagnetic interaction between elementary excitations in the NPs (localized surface plasmons)  and in the matrix (bound or unbound excitons). We derive a compact formula that generalizes Eq. (\ref{bare_eq1}) for this case. Its application to the situation where a sharp excitonic transition is superimposed on a broad LSPR band yields a Fano-type resonance. We also analyze the calculated LSPR lineshape in the case of CuO matrix where the interband transition is little affected by the electron-hole interaction and compare it with experimental results for the Au-CuO system.

\section{Renormalization of NP's polarizability in absorbing matrix}

\begin{figure}
\begin{center}
\includegraphics*[width=8cm]{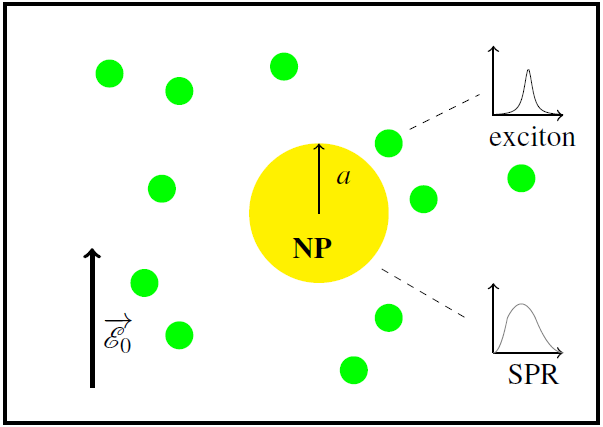}
\end{center}
\caption{Nanoparticle embedded in an  absorbing medium containing randomly dispersed entities (excitons) that are polarized in resonance with the applied EM field. }
\label{fig:scheme}
\end{figure}

Let us consider a metallic nanoparticle (NP) embedded in a medium that contains some resontantly polarizable entities randomly dispersed everywhere arround the NP (see Fig.\ref{fig:scheme}). These entities can be excitons or weakly interacting electron-hole pairs that can be created in the matrix if the incident EM field is in resonance with appropriate electronic transitions. They are responsible for absorption of the EM field in the matrix.
For clarity we shall call these polarizable entities by excitons and our purpose is to evaluate their influence on the NP's polarizability. 
If there were no excitons, NP would be described by the bare polarizability, in the electrostatic limit given by Eq. (\ref{bare_eq1}) with $\epsilon _h$ replaced by  $\epsilon_{\infty}\equiv \eta ^2 = const$, which is the (real) dielectric constant of the medium without excitons and $\eta $ is its refractive index. 

On the other hand, the medium containing excitons under illumination but free from NP inclusions has a dispersive dielectric function, $\epsilon (\omega )$, characteristic of a bulk semiconductor and its imaginary part,  $\epsilon ^{\prime \prime}(\omega )=\text {Im}\, \epsilon (\omega )$, represents light absorption owing to bound and unbound exciton states.\cite{Basu} This dielectric function is related to the exciton polarizability by an equation similar to the Clausius-Mossotti (CM) formula.\cite{Jackson} Before considering a NP, we shall derive this equation   by considering  the excitonic polarization  of the semiconductor medium that is polarized by an external EM field in resonance with the excitonic transition.
We shall assume that the external EM field has the electric component of the form,
 \begin{equation}\label{HN_MV_eq4}
\vec{{\cal E} _0}(\vec{r},t)=\vec{E_0} e^{i\vec q \vec{r}};\qquad \vec{E_0} \sim e^{-i\omega t},
\end{equation}
where $\vec{q}$ is the wave vector and $\omega =qc/\sqrt{\epsilon_{\infty}}$. \cite{Note1}
The polarization (i.e. the dipole moment per unit of volume) can be described by the following integral equation similar to the so called Lippmann-Schwinger equation adapted to the context of EM field scattering.\cite{Amorim2017} 
\begin{equation}\label{HN_MV_eq2}
\vec {P}(\vec{r})=\chi_e (\omega) \left[\vec{{\cal E} _0}+\int_{V'} \hat{T} ( \vec{r}- \vec{r ^\prime},\omega)  \vec{P}(\vec{r ^\prime}) d \vec{r^\prime}\right],
\end{equation}
where $\chi_e$ is the excitonic susceptibility. This equation is also analogous to the coupled dipole equations used to describe ensembles of polaizable NPs.\cite{Shalaev,Pereira2013,Vieaud2018}
In the last term of Eq. (\ref{HN_MV_eq2}), we should exclude the self-action, therefore $V'$ is the volume of the system excluding a small region arround $\vec{r}$, which is "occupied" by the exciton. Let us denote the radius of this spherical region  by $b$.
 In Eq. (\ref{HN_MV_eq2}) the dipole-dipole interaction tensor is:\cite{Landau-Lifshitz_II} 
\begin{equation}\label{HN_MV_eq5}
\hat {T}(\vec{R},\omega )= \hat{I} q^2 \frac{e^{iqR}}{R}+
\vec{\nabla} \otimes \vec{\nabla} (\frac{e^{iqR}}{R}),
\end{equation}
where $\hat{I}$ is the cartesian unit matrix, and $\otimes$ means the direct product. 

Equation (\ref{HN_MV_eq2}) can be solved by performing Fourier transform (see Appendix A) and it yields the polarization,
\begin{equation}\label{EP}
\vec {P}(\vec {r})=\frac{ \chi_e}{1-\frac{4 \pi}{3} \chi_e}\vec{{\cal E} _0}\;.
\end{equation}
Since the electric displacement vector is $\vec {D}=\epsilon_{\infty}(\vec{{\cal E} _0}+4\pi \vec {P})$, 
 from (\ref{EP})  follows the CM-type relation for the dielectric function,
 \begin{equation}\label{HN_MV_eq7}
\epsilon=\epsilon_{\infty} \frac{1+\frac{8 \pi}{3} \chi_e}{1-\frac{4 \pi}{3} \chi_e}.
\end{equation}

Now we shall insert one NP in the medium. The dipole moment of the NP can be written as
\begin{equation}\label{HN_MV_eq9}
\vec {p_0}=\alpha_0 \vec {E_0}+\int_{|\vec {r}|>a} \hat{T}(\vec {r},\omega )
\vec{P}(\vec{r}) d\vec{r},
\end{equation}
where the second term represents the field created by the excitonic polarization of the medium at the origin where the NP is placed.  Equation (\ref{HN_MV_eq2}) for the polarization  now has to include an additional term due to the field created by the NP:
\begin{equation}\label{HN_MV_eq10}
\vec{P}(\vec{r})=\chi_e \left[\vec{{\cal E} _0}+\int_{| \vec{r}- \vec{r^ \prime}|>b}  \hat{T} ( \vec{r}- \vec{r^ \prime},\omega)  \vec{P}(\vec{r^ \prime}) d \vec{r^ \prime}+ 
 \hat{T}(\hat{r},\omega)\vec{p_0}\right].
\end{equation}
Repeating the same steps as for pure matrix (see Appendix A) and Fourrier transforming the last term in (\ref{HN_MV_eq10}) forward and back, we obtain:
\begin{equation}\label{HN_MV_eq11}
\vec{P}(\vec{r})=\frac{\chi_e}{1-\frac{4\pi}{3}\chi_e}\left[\vec{{\cal E} _0}+
\hat{T}(\vec{r})\vec{p_0} \right]\;,
\end{equation}
where the argument $\omega $ was skipped for clarity.
Finally, substituting (\ref{HN_MV_eq11}) into (\ref{HN_MV_eq10}), we get: 
\begin{equation}\label{HN_MV_eq12}
\begin{array}{r@{}l}
  \vec {p_0}=\alpha_0 \left [ \vec{E_0}+\frac{\chi_e}{1-\frac{4\pi}{3}\chi_e}
\int_{|\vec{r}|>a} {\hat{T}(\vec{r})\vec{{\cal E} _0} d\vec{r}}\right. \\
  \left. +\frac{\chi_e}{1-\frac{4\pi}{3}\chi_e}  \int_{|\vec{r}|>a} \hat{T}(\vec{r})\left (\hat{T}(\vec{r})\vec{p_0} \right )
 d\vec{r} \right ]
\end{array}
\end{equation}
The second term in the brackets in Eq. (\ref{HN_MV_eq12}) vanishes in the limit  $\vec q \rightarrow 0$ by virtue of the angular integration, while that in the second line is evaluated to [see Appendix B, Eq. (\ref{A2_1})]:
$$
\text {2-nd line}=\alpha_0 \frac{\chi_e}{1-\frac{4\pi}{3}\chi_e} \frac{8 \pi}{3a^3} \vec {p_0}\;.
$$
Thus, we obtain:
\begin{equation}\label{HN_MV_eq14}
 \vec p_0=\left[\frac{\alpha_0}{1-\frac{8\pi}{3}\frac{\chi_e}{1-\frac{4\pi}{3}\chi_e}(\frac{\alpha_0}{a^3})   }\right] \vec{E_0}\, .
\end{equation}
The proportionality coefficient in (\ref{HN_MV_eq14}) may be called renormalized polarizability. 
We  can express the excitonic susceptibility,  $\chi_e$, in terms of the matrix dielectric function, $ \epsilon _h (\omega)=\epsilon_{\infty}+4\pi \chi _e (\omega)$, which yields:
\begin{equation}\label{HN_MV_eq15}
\alpha (\omega)
=\alpha_0(\omega)\left [{1-2\left (\frac{\alpha_0 (\omega)}{a^3}\right )\frac{\epsilon _h (\omega)-\epsilon_{\infty}}{\epsilon _h (\omega) +2\epsilon_{\infty}}}\right ]^{-1}\, .
\end{equation}
This is our main result: the NP's polarizability is renormalized because of the back action of the excitonic polarization in the semiconductor matrix, induced by the polarized NP. 
     
\section{Discussion and comparison with experiment}

The denominator of Eq. (\ref {HN_MV_eq15}) represents the effect of the EM wave dispersion and absorption in the host material, which is measured by $[\epsilon _h (\omega) - \epsilon_{\infty}]$.  In a non-absorbing matrix with $\epsilon _h =\epsilon_{\infty} = const$ we have $\alpha =\alpha _0$; also if the dispersion and absorption are small, they have little impact on the NP's polarizability.
However, if a sharp excitonic resonance overlaps with the surface plasmon resonance in the NP, the denominator of (\ref {HN_MV_eq15}) has a strong and non-trivial dependence upon the frequency in the vicinity of the overlap range. Indeed, in this spectral region the imaginary parts of $\alpha _0$ and $[\epsilon (\omega)-\epsilon_{\infty}]$  both are large, while the real parts may change sign and one may expect peaks and dips in the imaginary part of  $\alpha (\omega)$. This situation is characteristic of Fano-type resonances which occur when a discrete state (e.g. bound exciton) interferes with a continuum band of states (e.g. the relatively broad LSPR band).\cite{Fano,Lukyanchuk2010,Limonov2017,Bin2018}    

\begin{figure}
\begin{center}
\includegraphics*[width=16cm]{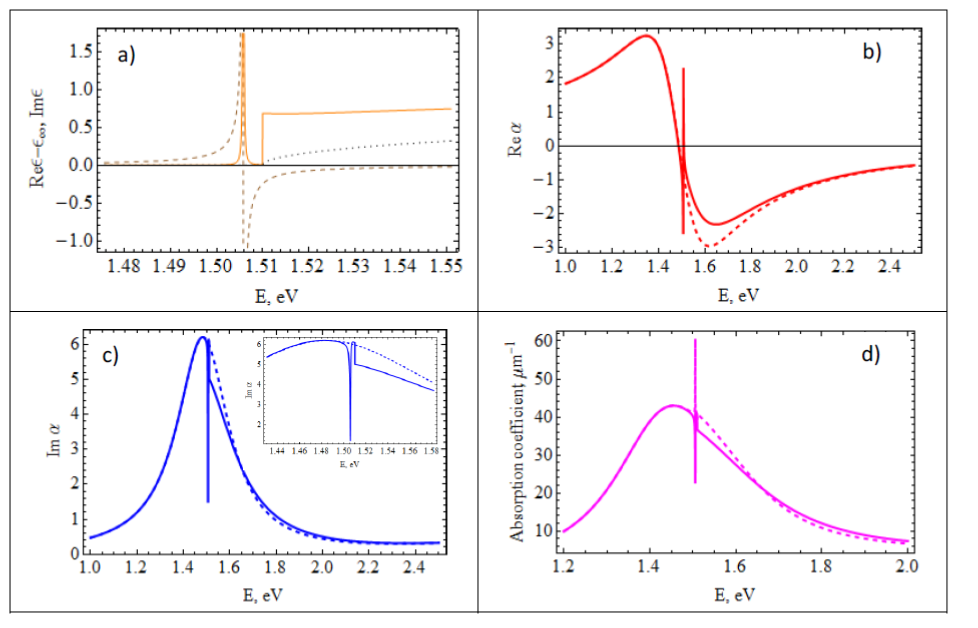}
\end{center}
\caption{(a): Dielectric function of GaAs including bound exciton and interband transitions (real part, full curve and imaginary part, dashed curve) [see Appendix C for details]; (b) and (c): Real ans imaginary parts of the NP polarizability, bare (dashed curves) and renormalized (full curves); (d) absorption coefficient of a composite material containing $f=0.1$ volume fraction of NPs, calculated using the modified Maxwell-Garnett approximation.\cite {VA96} }
\label{fig:GaAs}
\end{figure}

As an example, in Fig. \ref {fig:GaAs} we present the renormalized polarizability calculated for a gold NP embedded in GaAs. At low temperatures, the optical response of GaAs includes a sharp peak due to the bound exciton (in its ground state) and the interband absorption band, which is also modified by the exciton effect. Using the parametrization presented in Appendix C, the dielectric function is shown in Fig. \ref {fig:GaAs}a. The major effect in the renormalized polarizability comes from the bound exciton transition, it produces a sharp dip in the imaginary part of $\alpha (\omega)$ and a corresponding jump in its real part (Fig. \ref {fig:GaAs}b and c). The figure also presents the absorption coefficient, $A $, of a hypothetical composite film of GaAs containing gold inclusions; the Maxwell-Garnett approximation \cite{MG,VA96} was used for this purpose, which roughly leads to $A =(3\alpha /{a^3})f\sqrt {\epsilon_{\infty}} \omega /c $, where $f$ is the volume filling fraction of gold. The spectrum of the absorption coefficient presents the typical asymmetric lineshape (inset in Fig. \ref {fig:GaAs}d)   characteristic of Fano resonances that have been observed for different systems, see e.g. Ref.\cite{Limonov2017} 

\begin{figure}
\begin{center}
\includegraphics*[width=14cm]{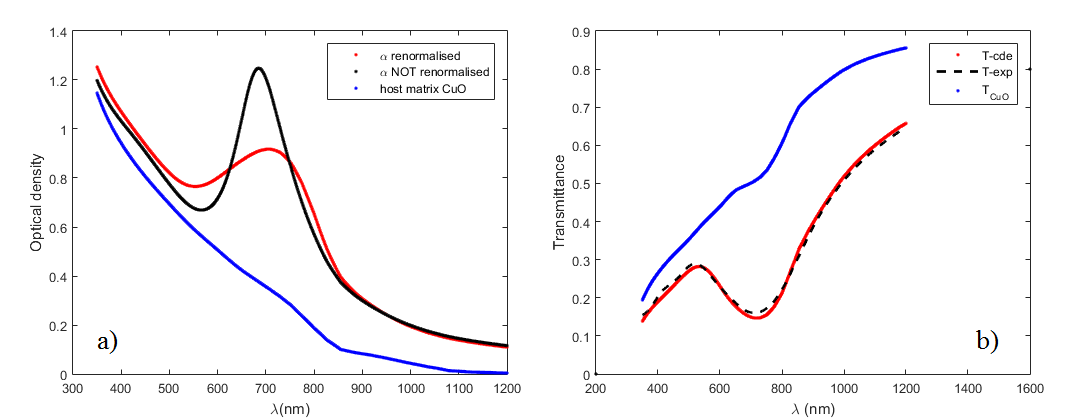}
\end{center}
\caption{(a) Optical density calculated using the renormalized polarizability (red curve) and the bare one (black curve); also shown is  an experimental spectrum of host matrix (blue line).
(b) Transmittance of the composite Au/CuO film described in the text, experimental (black dashed curve) and calculated numerically (red curve) using Eqs. (\ref{HN_MV_eq16}) and  (\ref {HN_MV_eq38}); also shown  is the transmittance of the host matrix CuO (blue line).
The parameters of gold were taken from Ref. \cite{Au}} 
\label{fig3}
\end{figure}

Interband transitions in the matrix material also influence the renormalized polarizability and the absorption coefficient, although in a less drastic way. In order to look closer at this effect, we considered the situation of higher temperatures where bound exciton states are dissolved and the interband absorption in the semiconductor follows the simple parabolic law.\cite{Basu}  In the absence of the bound exciton contribution, we shall take the background dielectric constant as the square of the refractive index, $\epsilon_{\infty}=\eta^2$, and $\epsilon =\eta^2+i\epsilon ^{\prime \prime }(\omega) $. Therefore equation (\ref{HN_MV_eq15}) becomes:
\begin{equation}
\alpha  (\omega)=\frac{\alpha_0  (\omega)}{1-2(\frac{\alpha_0 (\omega)}{a^3})\frac{i\epsilon _h ^{\prime \prime }(\omega)}{3\eta^2+i\epsilon _h ^{\prime \prime }(\omega)}}.
\label{HN_MV_eq16}
\end{equation}
Here the imaginary part of the host dielectric function just increases monotonically with $\omega $ above the absorption edge,
$$
\epsilon _h ^{\prime \prime }(\omega) \propto \frac {\sqrt {\hbar \omega-E_g}}{\omega ^2}\, ,
$$
and the situation, in a sense, is the opposite compared to the previous case, i.e. LSPR is superimposed on a continuum of interband transitions. The effect can already be noticed Fig. \ref{fig:GaAs}d (above $E_g$) but it is illustrated better in Fig. \ref{fig3}a for the case of Au NPs in CuO matrix. 
The optical density of such a composite material was calculated using its effective dielectric function obtained by solving coupled dipole equations for an ensemble of identical NPs randomly distributed over sites of a mesh with the occupation probability that guarantees the required volume fraction of the metal (see Refs. \cite{Pereira2013,Pereira2016}  for details of this procedure).
In this figure we observe an asymmetric broadening of the LSPR band caused by its superposition with the parabolic absorption band of the semiconductor but the line shape is not like in the usual case of Fano-type resonances.

Next we present the results obtained from our experiment with a film (about 40 nm in thickness) with approximadetly 15\% of gold (by volume fraction, determined by the Rutherford Backscattering Spectrometry technique), dispersed in a CuO matrix. The nanocomposite film was deposited by magnetron sputtering and then annealed to stimulate the NP growth. The average size of Au NPs in the sample, estimated by X-Ray Diffraction peak fitting was $\approx  \,$12 nm. More details concerning the deposition and characterization of Au/CuO films can be found in Ref.\cite{Proenca2017}  
The experimental transmittance spectrum is very  well reproduced by the theory using the NPs' parameters evaluated experimentally by independent techniques . 
This fit is much better than any one which we could obtain using the bare polarizability (\ref{bare_eq1}) and eventually prescribing an imaginary prescribing and imaginary part to $\epsilon _h$. We emphasize that such a good fit could not be obtained even adjusting, say, the NPs' filling fraction. Therefore we are convinced that it demonstrates the importance of the renormalization effect in matrices where interband light absorption, overlapping with the LSPR, takes place.

\section{Conclusion}
In summary, we have shown that the optical properties of a nanocomposite system composed by plasmonic nanoparticles (e. g. Au nanospheres) embedded in a semiconductor matrix (e.g. CuO) results from the coupling between localized surface plasmons and excitons or free-carrier interband transitions. In the latter case, characteristic of CuO at room temperature (because of the small exciton binding energy), the use of the renormalized polarizability derived in this work [Eq. \ref{HN_MV_eq15}] permits to obtain a very good fit to the experimentally measured LSPR lineshape.  

For the theoretically considered situation where a sharp exciton transition exists in the matrix, spectrally overlapping with the LSPR band, the exciton interaction with localized surface plasmons results in a Fano-type resonance with its characteristic non-Lorentzian lineshape (Fig.  \ref{fig:GaAs}). A steep spectral feature (the Fano resonance) combined with a broad plasmonic band is intersting from the point of view of complementarity of spectral shift assays related to refractive index changes and SERS molecular identification.\cite{Anker2008} Indeed, such a dual-sensor idea based on Fano resonances originating from the interaction between a (broad) dipole mode and a (narrow) quadrupole one, produced by two different plasmonic components of the nanostructure (e.g. spheres and rings) put in a contact.\cite{Bin2018} A tunable Fano resonance in a resonator combining metallic NPs with a dielectric microcavity was demonstrated in Ref. \cite{Gu2016} It is also interesting to explore a structure combining plasmonic NPs with a two-dimensional semiconductor such as transition metal dichalcogenides,\cite{Abid2017,Liu2018} which are known for their  strong excitonic transitions.\cite{Wang2018} 
Such nanostructures promise the advantages of enhanced sensitivity and a better figure of merit compared to dipole-based localized surface plasmon resonance (LSPR) sensors.\cite{Bin2018}


\section*{Acknowledgements}
Financial  support  from  the Portuguese Foundation for Science and Technology (FCT) in the framework of the Strategic Financing UID/FIS/04650/2013 is acknowledged. This research was also partially sponsored in the framework of  FCT Project 9471 - RIDTI, co-funded by FEDER (Ref. PTDC/FIS-NAN/1154/2014). Joel Borges thanks FCT for Postdoctorate Grant (SFRH/BPD/117010/2016). 




\newpage

\bibliographystyle{apsrev}
\bibliography{Plasmonic_NPs}

\begin{thebibliography}{41}
\expandafter\ifx\csname natexlab\endcsname\relax\def\natexlab#1{#1}\fi
\expandafter\ifx\csname bibnamefont\endcsname\relax
  \def\bibnamefont#1{#1}\fi
\expandafter\ifx\csname bibfnamefont\endcsname\relax
  \def\bibfnamefont#1{#1}\fi
\expandafter\ifx\csname citenamefont\endcsname\relax
  \def\citenamefont#1{#1}\fi
\expandafter\ifx\csname url\endcsname\relax
  \def\url#1{\texttt{#1}}\fi
\expandafter\ifx\csname urlprefix\endcsname\relax\def\urlprefix{URL }\fi
\providecommand{\bibinfo}[2]{#2}
\providecommand{\eprint}[2][]{\url{#2}}

\bibitem[{\citenamefont{Bohren and Huffman}(1998)}]{Bohren-Huffman}
\bibinfo{author}{\bibfnamefont{C.~F.} \bibnamefont{Bohren}} \bibnamefont{and}
  \bibinfo{author}{\bibfnamefont{D.~R.} \bibnamefont{Huffman}},
  \emph{\bibinfo{title}{Absorption and scattering of light by small particles}}
  (\bibinfo{publisher}{Wiley, NY}, \bibinfo{year}{1998}).

\bibitem[{\citenamefont{Stockman}(2011)}]{Stockman2011}
\bibinfo{author}{\bibfnamefont{M.}~\bibnamefont{Stockman}},
  \bibinfo{journal}{Physics Today} \textbf{\bibinfo{volume}{64}},
  \bibinfo{pages}{39} (\bibinfo{year}{2011}).

\bibitem[{\citenamefont{Garcia~de Abajo}(2007)}]{GarciadeAbajo2007}
\bibinfo{author}{\bibfnamefont{F.~J.} \bibnamefont{Garcia~de Abajo}},
  \bibinfo{journal}{Rev. Mod. Phys.} \textbf{\bibinfo{volume}{79}},
  \bibinfo{pages}{1267 } (\bibinfo{year}{2007}).

\bibitem[{\citenamefont{Lachaine et~al.}(2016)\citenamefont{Lachaine, Boulais,
  Rioux, Boutopoulos, and Meunier}}]{Lachaine2016}
\bibinfo{author}{\bibfnamefont{R.}~\bibnamefont{Lachaine}},
  \bibinfo{author}{\bibfnamefont{E.}~\bibnamefont{Boulais}},
  \bibinfo{author}{\bibfnamefont{D.}~\bibnamefont{Rioux}},
  \bibinfo{author}{\bibfnamefont{C.}~\bibnamefont{Boutopoulos}},
  \bibnamefont{and} \bibinfo{author}{\bibfnamefont{M.}~\bibnamefont{Meunier}},
  \bibinfo{journal}{ACS Photonics} \textbf{\bibinfo{volume}{3}},
  \bibinfo{pages}{2158–2169} (\bibinfo{year}{2016}).

\bibitem[{\citenamefont{Amorim et~al.}(2017)\citenamefont{Amorim, Gonçalves,
  Vasilevskiy, and Peres}}]{Amorim2017}
\bibinfo{author}{\bibfnamefont{B.}~\bibnamefont{Amorim}},
  \bibinfo{author}{\bibfnamefont{P.~A.~D.} \bibnamefont{Gonçalves}},
  \bibinfo{author}{\bibfnamefont{M.~I.} \bibnamefont{Vasilevskiy}},
  \bibnamefont{and} \bibinfo{author}{\bibfnamefont{N.~M.~R.}
  \bibnamefont{Peres}}, \bibinfo{journal}{Appl. Sci.}
  \textbf{\bibinfo{volume}{7}}, \bibinfo{pages}{1158} (\bibinfo{year}{2017}).

\bibitem[{\citenamefont{Abid et~al.}(2017)\citenamefont{Abid, Chen, Yuan,
  Bohloul, Najmaei, Avendano, P\'echou, Mlayah, and Low}}]{Abid2017}
\bibinfo{author}{\bibfnamefont{I.}~\bibnamefont{Abid}},
  \bibinfo{author}{\bibfnamefont{W.}~\bibnamefont{Chen}},
  \bibinfo{author}{\bibfnamefont{J.}~\bibnamefont{Yuan}},
  \bibinfo{author}{\bibfnamefont{A.}~\bibnamefont{Bohloul}},
  \bibinfo{author}{\bibfnamefont{S.}~\bibnamefont{Najmaei}},
  \bibinfo{author}{\bibfnamefont{C.}~\bibnamefont{Avendano}},
  \bibinfo{author}{\bibfnamefont{R.}~\bibnamefont{P\'echou}},
  \bibinfo{author}{\bibfnamefont{A.}~\bibnamefont{Mlayah}}, \bibnamefont{and}
  \bibinfo{author}{\bibfnamefont{J.}~\bibnamefont{Low}}, \bibinfo{journal}{ACS
  Photonics} \textbf{\bibinfo{volume}{4}}, \bibinfo{pages}{1653–1660}
  (\bibinfo{year}{2017}).

\bibitem[{\citenamefont{Anker et~al.}(2008)\citenamefont{Anker, Hall, Lyandres,
  Shah, Zhao, and Duyne}}]{Anker2008}
\bibinfo{author}{\bibfnamefont{J.~N.} \bibnamefont{Anker}},
  \bibinfo{author}{\bibfnamefont{W.~P.} \bibnamefont{Hall}},
  \bibinfo{author}{\bibfnamefont{O.}~\bibnamefont{Lyandres}},
  \bibinfo{author}{\bibfnamefont{N.~C.} \bibnamefont{Shah}},
  \bibinfo{author}{\bibfnamefont{J.}~\bibnamefont{Zhao}}, \bibnamefont{and}
  \bibinfo{author}{\bibfnamefont{R.~P.~V.} \bibnamefont{Duyne}},
  \bibinfo{journal}{Nature Materials} \textbf{\bibinfo{volume}{7}},
  \bibinfo{pages}{242} (\bibinfo{year}{2008}).

\bibitem[{\citenamefont{Cappi et~al.}(2015)\citenamefont{Cappi, Spiga, Moncada,
  Ferretti, Beyeler, Bianchessi, Decosterd, Buclin, and Giuducci}}]{Cappi2015}
\bibinfo{author}{\bibfnamefont{G.}~\bibnamefont{Cappi}},
  \bibinfo{author}{\bibfnamefont{F.~M.} \bibnamefont{Spiga}},
  \bibinfo{author}{\bibfnamefont{Y.}~\bibnamefont{Moncada}},
  \bibinfo{author}{\bibfnamefont{A.}~\bibnamefont{Ferretti}},
  \bibinfo{author}{\bibfnamefont{M.}~\bibnamefont{Beyeler}},
  \bibinfo{author}{\bibfnamefont{M.}~\bibnamefont{Bianchessi}},
  \bibinfo{author}{\bibfnamefont{I.}~\bibnamefont{Decosterd}},
  \bibinfo{author}{\bibfnamefont{T.}~\bibnamefont{Buclin}}, \bibnamefont{and}
  \bibinfo{author}{\bibfnamefont{C.}~\bibnamefont{Giuducci}},
  \bibinfo{journal}{Anal. Chem.} \textbf{\bibinfo{volume}{87}},
  \bibinfo{pages}{5278} (\bibinfo{year}{2015}).

\bibitem[{\citenamefont{Godselahi et~al.}(2011)\citenamefont{Godselahi,
  Zahrabi, Saani, and Vesaghi}}]{Ghod2011}
\bibinfo{author}{\bibfnamefont{T.}~\bibnamefont{Godselahi}},
  \bibinfo{author}{\bibfnamefont{H.}~\bibnamefont{Zahrabi}},
  \bibinfo{author}{\bibfnamefont{M.~H.} \bibnamefont{Saani}}, \bibnamefont{and}
  \bibinfo{author}{\bibfnamefont{M.~A.} \bibnamefont{Vesaghi}},
  \bibinfo{journal}{J. Phys. Chem. C} \textbf{\bibinfo{volume}{115}},
  \bibinfo{pages}{22126} (\bibinfo{year}{2011}).

\bibitem[{\citenamefont{Novotny and van Hulst}(2011)}]{Novotny2011}
\bibinfo{author}{\bibfnamefont{L.}~\bibnamefont{Novotny}} \bibnamefont{and}
  \bibinfo{author}{\bibfnamefont{N.}~\bibnamefont{van Hulst}},
  \bibinfo{journal}{Nature Photonics} \textbf{\bibinfo{volume}{5}},
  \bibinfo{pages}{83} (\bibinfo{year}{2011}).

\bibitem[{\citenamefont{Schreiber et~al.}(2014)\citenamefont{Schreiber, Do,
  Roller, Zhang, Schaller, Nickels, Feldmann, and Liedl}}]{Schreiber2014}
\bibinfo{author}{\bibfnamefont{R.}~\bibnamefont{Schreiber}},
  \bibinfo{author}{\bibfnamefont{J.}~\bibnamefont{Do}},
  \bibinfo{author}{\bibfnamefont{E.-M.} \bibnamefont{Roller}},
  \bibinfo{author}{\bibfnamefont{T.}~\bibnamefont{Zhang}},
  \bibinfo{author}{\bibfnamefont{V.~J.} \bibnamefont{Schaller}},
  \bibinfo{author}{\bibfnamefont{P.~C.} \bibnamefont{Nickels}},
  \bibinfo{author}{\bibfnamefont{J.}~\bibnamefont{Feldmann}}, \bibnamefont{and}
  \bibinfo{author}{\bibfnamefont{T.}~\bibnamefont{Liedl}},
  \bibinfo{journal}{Nature Nanotechnology} \textbf{\bibinfo{volume}{9}},
  \bibinfo{pages}{74} (\bibinfo{year}{2014}).

\bibitem[{\citenamefont{Chumanov et~al.}(1995)\citenamefont{Chumanov, Sokolov,
  Gregory, and Cotton}}]{Chumanov}
\bibinfo{author}{\bibfnamefont{G.}~\bibnamefont{Chumanov}},
  \bibinfo{author}{\bibfnamefont{K.}~\bibnamefont{Sokolov}},
  \bibinfo{author}{\bibfnamefont{B.~W.} \bibnamefont{Gregory}},
  \bibnamefont{and} \bibinfo{author}{\bibfnamefont{T.~M.}
  \bibnamefont{Cotton}}, \bibinfo{journal}{J. Phys. Chem.}
  \textbf{\bibinfo{volume}{99}}, \bibinfo{pages}{9466 } (\bibinfo{year}{1995}).

\bibitem[{\citenamefont{Basko et~al.}(2000)\citenamefont{Basko, Agranovich,
  Bassani, and Rocca}}]{Basko2000}
\bibinfo{author}{\bibfnamefont{D.~M.} \bibnamefont{Basko}},
  \bibinfo{author}{\bibfnamefont{V.~M.} \bibnamefont{Agranovich}},
  \bibinfo{author}{\bibfnamefont{F.}~\bibnamefont{Bassani}}, \bibnamefont{and}
  \bibinfo{author}{\bibfnamefont{G.~C.~L.} \bibnamefont{Rocca}},
  \bibinfo{journal}{Eur. Phys. J. B} \textbf{\bibinfo{volume}{13}},
  \bibinfo{pages}{653} (\bibinfo{year}{2000}).

\bibitem[{\citenamefont{Carnegie et~al.}(2017)\citenamefont{Carnegie,
  Chikkaraddy, Benz, de~Nijs, Deacon, Horton, Wang, Readman, Barrow, Scherman
  et~al.}}]{Carnegie2017}
\bibinfo{author}{\bibfnamefont{C.}~\bibnamefont{Carnegie}},
  \bibinfo{author}{\bibfnamefont{R.}~\bibnamefont{Chikkaraddy}},
  \bibinfo{author}{\bibfnamefont{F.}~\bibnamefont{Benz}},
  \bibinfo{author}{\bibfnamefont{B.}~\bibnamefont{de~Nijs}},
  \bibinfo{author}{\bibfnamefont{W.~M.} \bibnamefont{Deacon}},
  \bibinfo{author}{\bibfnamefont{M.}~\bibnamefont{Horton}},
  \bibinfo{author}{\bibfnamefont{W.}~\bibnamefont{Wang}},
  \bibinfo{author}{\bibfnamefont{C.}~\bibnamefont{Readman}},
  \bibinfo{author}{\bibfnamefont{S.~J.} \bibnamefont{Barrow}},
  \bibinfo{author}{\bibfnamefont{O.~A.} \bibnamefont{Scherman}},
  \bibnamefont{et~al.}, \bibinfo{journal}{ACS Photonics}
  \textbf{\bibinfo{volume}{4}}, \bibinfo{pages}{2681–2686}
  (\bibinfo{year}{2017}).

\bibitem[{\citenamefont{Qiao et~al.}(2011)\citenamefont{Qiao, Wang, Zuo, Ye,
  Qian, Chen, and He}}]{Quiao2011}
\bibinfo{author}{\bibfnamefont{L.}~\bibnamefont{Qiao}},
  \bibinfo{author}{\bibfnamefont{D.}~\bibnamefont{Wang}},
  \bibinfo{author}{\bibfnamefont{L.}~\bibnamefont{Zuo}},
  \bibinfo{author}{\bibfnamefont{Y.}~\bibnamefont{Ye}},
  \bibinfo{author}{\bibfnamefont{J.}~\bibnamefont{Qian}},
  \bibinfo{author}{\bibfnamefont{H.}~\bibnamefont{Chen}}, \bibnamefont{and}
  \bibinfo{author}{\bibfnamefont{S.}~\bibnamefont{He}},
  \bibinfo{journal}{Applied Energy} \textbf{\bibinfo{volume}{88}},
  \bibinfo{pages}{848–852} (\bibinfo{year}{2011}).

\bibitem[{\citenamefont{Hill et~al.}(2014)\citenamefont{Hill, Kozek, Hucknall,
  Smith, and Chilkoti}}]{Hill2014}
\bibinfo{author}{\bibfnamefont{R.~T.} \bibnamefont{Hill}},
  \bibinfo{author}{\bibfnamefont{K.~M.} \bibnamefont{Kozek}},
  \bibinfo{author}{\bibfnamefont{A.}~\bibnamefont{Hucknall}},
  \bibinfo{author}{\bibfnamefont{D.~R.} \bibnamefont{Smith}}, \bibnamefont{and}
  \bibinfo{author}{\bibfnamefont{A.}~\bibnamefont{Chilkoti}},
  \bibinfo{journal}{ACS Photonics} \textbf{\bibinfo{volume}{1}},
  \bibinfo{pages}{974–984} (\bibinfo{year}{2014}).

\bibitem[{\citenamefont{Blackman}(2008)}]{Blackman}
\bibinfo{author}{\bibfnamefont{J.}~\bibnamefont{Blackman}},
  \emph{\bibinfo{title}{Metallic Nanoparticles}} (\bibinfo{publisher}{Elsevier,
  Amsterdam}, \bibinfo{year}{2008}).

\bibitem[{\citenamefont{Maxwell-Garnett}(1904)}]{MG}
\bibinfo{author}{\bibfnamefont{J.~C.} \bibnamefont{Maxwell-Garnett}},
  \bibinfo{journal}{Philos. Trans. R. Soc. London}
  \textbf{\bibinfo{volume}{203}}, \bibinfo{pages}{385} (\bibinfo{year}{1904}).

\bibitem[{\citenamefont{Shalaev}(2000)}]{Shalaev}
\bibinfo{author}{\bibfnamefont{V.~M.} \bibnamefont{Shalaev}},
  \emph{\bibinfo{title}{Non-linear Optics of Random Media}}
  (\bibinfo{publisher}{Clarendon: Oxford}, \bibinfo{year}{2000}).

\bibitem[{\citenamefont{Pereira et~al.}(2013)\citenamefont{Pereira, Pereira,
  Smirnov, and Vasilevskiy}}]{Pereira2013}
\bibinfo{author}{\bibfnamefont{R.~M.~S.} \bibnamefont{Pereira}},
  \bibinfo{author}{\bibfnamefont{P.}~\bibnamefont{Pereira}},
  \bibinfo{author}{\bibfnamefont{G.}~\bibnamefont{Smirnov}}, \bibnamefont{and}
  \bibinfo{author}{\bibfnamefont{M.~I.} \bibnamefont{Vasilevskiy}},
  \bibinfo{journal}{Europhysics Letters} \textbf{\bibinfo{volume}{102}},
  \bibinfo{pages}{67001} (\bibinfo{year}{2013}).

\bibitem[{\citenamefont{Vieaud et~al.}(2018)\citenamefont{Vieaud, Gao, Cane,
  Stchakovsky, Naciri, Ariga, Oda, Pouget, and Battie}}]{Vieaud2018}
\bibinfo{author}{\bibfnamefont{J.}~\bibnamefont{Vieaud}},
  \bibinfo{author}{\bibfnamefont{J.}~\bibnamefont{Gao}},
  \bibinfo{author}{\bibfnamefont{J.}~\bibnamefont{Cane}},
  \bibinfo{author}{\bibfnamefont{M.}~\bibnamefont{Stchakovsky}},
  \bibinfo{author}{\bibfnamefont{A.~E.} \bibnamefont{Naciri}},
  \bibinfo{author}{\bibfnamefont{K.}~\bibnamefont{Ariga}},
  \bibinfo{author}{\bibfnamefont{R.}~\bibnamefont{Oda}},
  \bibinfo{author}{\bibfnamefont{E.}~\bibnamefont{Pouget}}, \bibnamefont{and}
  \bibinfo{author}{\bibfnamefont{Y.}~\bibnamefont{Battie}},
  \bibinfo{journal}{J. Phys. Chem. C} \textbf{\bibinfo{volume}{122}},
  \bibinfo{pages}{11973−11984} (\bibinfo{year}{2018}).

\bibitem[{\citenamefont{Chen and Liedberg}(2014)}]{Chen2014}
\bibinfo{author}{\bibfnamefont{P.}~\bibnamefont{Chen}} \bibnamefont{and}
  \bibinfo{author}{\bibfnamefont{B.}~\bibnamefont{Liedberg}},
  \bibinfo{journal}{Anal. Chem.} \textbf{\bibinfo{volume}{86}},
  \bibinfo{pages}{7399} (\bibinfo{year}{2014}).

\bibitem[{\citenamefont{Chu et~al.}(2010)\citenamefont{Chu, Banaee, and
  Crozier}}]{Chu2010}
\bibinfo{author}{\bibfnamefont{Y.}~\bibnamefont{Chu}},
  \bibinfo{author}{\bibfnamefont{M.~G.} \bibnamefont{Banaee}},
  \bibnamefont{and} \bibinfo{author}{\bibfnamefont{K.~B.}
  \bibnamefont{Crozier}}, \bibinfo{journal}{ACS Nano}
  \textbf{\bibinfo{volume}{4}}, \bibinfo{pages}{2804–2810}
  (\bibinfo{year}{2010}).

\bibitem[{\citenamefont{Fano}(1961)}]{Fano}
\bibinfo{author}{\bibfnamefont{U.}~\bibnamefont{Fano}}, \bibinfo{journal}{Phys.
  Rev.} \textbf{\bibinfo{volume}{124}}, \bibinfo{pages}{1866}
  (\bibinfo{year}{1961}).

\bibitem[{\citenamefont{Lukyanchuk et~al.}(2010)\citenamefont{Lukyanchuk,
  Zheludev, Maier, Halas, Nordlander, Giessen, and Chong}}]{Lukyanchuk2010}
\bibinfo{author}{\bibfnamefont{B.}~\bibnamefont{Lukyanchuk}},
  \bibinfo{author}{\bibfnamefont{N.~I.} \bibnamefont{Zheludev}},
  \bibinfo{author}{\bibfnamefont{S.~A.} \bibnamefont{Maier}},
  \bibinfo{author}{\bibfnamefont{N.~J.} \bibnamefont{Halas}},
  \bibinfo{author}{\bibfnamefont{P.}~\bibnamefont{Nordlander}},
  \bibinfo{author}{\bibfnamefont{H.}~\bibnamefont{Giessen}}, \bibnamefont{and}
  \bibinfo{author}{\bibfnamefont{C.~T.} \bibnamefont{Chong}},
  \bibinfo{journal}{Nature Materials} \textbf{\bibinfo{volume}{9}},
  \bibinfo{pages}{707} (\bibinfo{year}{2010}).

\bibitem[{Not({\natexlab{a}})}]{Note2}
\bibinfo{note}{In this case the Fano resonance results from spectral overlap of
  dipole and quadrupole modes in the spheres.}

\bibitem[{\citenamefont{Proenca et~al.}(2018)\citenamefont{Proenca, Borges,
  Rodrigues, Domingues, Dias, Trigueiro, Bundaleski, Teodoro, and
  Vaz}}]{Proenca2017}
\bibinfo{author}{\bibfnamefont{M.}~\bibnamefont{Proenca}},
  \bibinfo{author}{\bibfnamefont{J.}~\bibnamefont{Borges}},
  \bibinfo{author}{\bibfnamefont{M.~S.} \bibnamefont{Rodrigues}},
  \bibinfo{author}{\bibfnamefont{R.~P.} \bibnamefont{Domingues}},
  \bibinfo{author}{\bibfnamefont{J.~P.} \bibnamefont{Dias}},
  \bibinfo{author}{\bibfnamefont{J.}~\bibnamefont{Trigueiro}},
  \bibinfo{author}{\bibfnamefont{N.}~\bibnamefont{Bundaleski}},
  \bibinfo{author}{\bibfnamefont{O.~M. N.~D.} \bibnamefont{Teodoro}},
  \bibnamefont{and} \bibinfo{author}{\bibfnamefont{F.}~\bibnamefont{Vaz}},
  \bibinfo{journal}{Surf. Coatings Technol.} \textbf{\bibinfo{volume}{343}},
  \bibinfo{pages}{178–185} (\bibinfo{year}{2018}).

\bibitem[{\citenamefont{Ribbins and Roos}(1997)}]{CuO}
\bibinfo{author}{\bibfnamefont{C.~G.} \bibnamefont{Ribbins}} \bibnamefont{and}
  \bibinfo{author}{\bibfnamefont{A.}~\bibnamefont{Roos}},
  \emph{\bibinfo{title}{In: Handbook on Optical Constants in Solids}}
  (\bibinfo{publisher}{Elsevier}, \bibinfo{year}{1997}).

\bibitem[{\citenamefont{Basu}(1997)}]{Basu}
\bibinfo{author}{\bibfnamefont{P.~K.} \bibnamefont{Basu}},
  \emph{\bibinfo{title}{Theory of Optical Processes in Semiconductors}}
  (\bibinfo{publisher}{Springer, Berlin}, \bibinfo{year}{1997}).

\bibitem[{\citenamefont{Jackson}(1998)}]{Jackson}
\bibinfo{author}{\bibfnamefont{J.~D.} \bibnamefont{Jackson}},
  \emph{\bibinfo{title}{Classical Electrodynamics}} (\bibinfo{publisher}{J.
  Wiley, New York}, \bibinfo{year}{1998}).

\bibitem[{Not({\natexlab{b}})}]{Note1}
\bibinfo{note}{Later we shall take the electrostatic limit, $q\rightarrow 0$.}

\bibitem[{\citenamefont{Landau and Lifshitz}(1971)}]{Landau-Lifshitz_II}
\bibinfo{author}{\bibfnamefont{L.~D.} \bibnamefont{Landau}} \bibnamefont{and}
  \bibinfo{author}{\bibfnamefont{E.~M.} \bibnamefont{Lifshitz}},
  \emph{\bibinfo{title}{The Classical Theory of Fields: Volume 2 (Course of
  Theoretical Physics)}} (\bibinfo{publisher}{Pergamon, Oxford},
  \bibinfo{year}{1971}).

\bibitem[{\citenamefont{Limonov et~al.}(2017)\citenamefont{Limonov, Rybin,
  Poddubny, and Kivshar}}]{Limonov2017}
\bibinfo{author}{\bibfnamefont{M.~F.} \bibnamefont{Limonov}},
  \bibinfo{author}{\bibfnamefont{M.~V.} \bibnamefont{Rybin}},
  \bibinfo{author}{\bibfnamefont{A.~N.} \bibnamefont{Poddubny}},
  \bibnamefont{and} \bibinfo{author}{\bibfnamefont{Y.~S.}
  \bibnamefont{Kivshar}}, \bibinfo{journal}{Nature Photonics}
  \textbf{\bibinfo{volume}{11}}, \bibinfo{pages}{543} (\bibinfo{year}{2017}).

\bibitem[{\citenamefont{Ai et~al.}(2018)\citenamefont{Ai, Song, Bradley, and
  Zhao}}]{Bin2018}
\bibinfo{author}{\bibfnamefont{B.}~\bibnamefont{Ai}},
  \bibinfo{author}{\bibfnamefont{C.}~\bibnamefont{Song}},
  \bibinfo{author}{\bibfnamefont{L.}~\bibnamefont{Bradley}}, \bibnamefont{and}
  \bibinfo{author}{\bibfnamefont{Y.}~\bibnamefont{Zhao}}, \bibinfo{journal}{J.
  Phys. Chem. C} \textbf{\bibinfo{volume}{0}}, \bibinfo{pages}{0}
  (\bibinfo{year}{2018}).

\bibitem[{\citenamefont{Vasilevskiy and Anda}(1996)}]{VA96}
\bibinfo{author}{\bibfnamefont{M.~I.} \bibnamefont{Vasilevskiy}}
  \bibnamefont{and} \bibinfo{author}{\bibfnamefont{E.~V.} \bibnamefont{Anda}},
  \bibinfo{journal}{Phys. Rev. B} \textbf{\bibinfo{volume}{54}},
  \bibinfo{pages}{5844} (\bibinfo{year}{1996}).

\bibitem[{\citenamefont{Borges et~al.}(2016)\citenamefont{Borges, Pereira,
  Rodrigues, Kubart, Sathyanath, Leifer, Cavaleiro, Polcar, Vasilevskiy, and
  Vaz}}]{Pereira2016}
\bibinfo{author}{\bibfnamefont{J.}~\bibnamefont{Borges}},
  \bibinfo{author}{\bibfnamefont{R.~M.~S.} \bibnamefont{Pereira}},
  \bibinfo{author}{\bibfnamefont{M.~S.} \bibnamefont{Rodrigues}},
  \bibinfo{author}{\bibfnamefont{T.}~\bibnamefont{Kubart}},
  \bibinfo{author}{\bibfnamefont{S.}~\bibnamefont{Sathyanath}},
  \bibinfo{author}{\bibfnamefont{K.}~\bibnamefont{Leifer}},
  \bibinfo{author}{\bibfnamefont{A.}~\bibnamefont{Cavaleiro}},
  \bibinfo{author}{\bibfnamefont{T.}~\bibnamefont{Polcar}},
  \bibinfo{author}{\bibfnamefont{M.~I.} \bibnamefont{Vasilevskiy}},
  \bibnamefont{and} \bibinfo{author}{\bibfnamefont{F.}~\bibnamefont{Vaz}},
  \bibinfo{journal}{J. Phys. Chem. C} \textbf{\bibinfo{volume}{120}},
  \bibinfo{pages}{16931} (\bibinfo{year}{2016}).

\bibitem[{\citenamefont{Etchegoin et~al.}(2007)\citenamefont{Etchegoin, Ru, and
  Meyer}}]{Au}
\bibinfo{author}{\bibfnamefont{P.~G.} \bibnamefont{Etchegoin}},
  \bibinfo{author}{\bibfnamefont{E.~L.} \bibnamefont{Ru}}, \bibnamefont{and}
  \bibinfo{author}{\bibfnamefont{M.}~\bibnamefont{Meyer}}, \bibinfo{journal}{J.
  Chem. Phys.} \textbf{\bibinfo{volume}{127}}, \bibinfo{pages}{189901}
  (\bibinfo{year}{2007}).

\bibitem[{\citenamefont{Gu et~al.}(2016)\citenamefont{Gu, Wan, Wu, Chen, and
  Wang}}]{Gu2016}
\bibinfo{author}{\bibfnamefont{P.}~\bibnamefont{Gu}},
  \bibinfo{author}{\bibfnamefont{M.}~\bibnamefont{Wan}},
  \bibinfo{author}{\bibfnamefont{W.}~\bibnamefont{Wu}},
  \bibinfo{author}{\bibfnamefont{Z.}~\bibnamefont{Chen}}, \bibnamefont{and}
  \bibinfo{author}{\bibfnamefont{Z.}~\bibnamefont{Wang}},
  \bibinfo{journal}{Nanoscale} \textbf{\bibinfo{volume}{8}},
  \bibinfo{pages}{103588–10363} (\bibinfo{year}{2016}).

\bibitem[{\citenamefont{Liu et~al.}(2018)\citenamefont{Liu, Wang, Naylor, Lee,
  Zheng, Liu, Johnson, Pan, and Agarwal}}]{Liu2018}
\bibinfo{author}{\bibfnamefont{W.}~\bibnamefont{Liu}},
  \bibinfo{author}{\bibfnamefont{Y.}~\bibnamefont{Wang}},
  \bibinfo{author}{\bibfnamefont{C.}~\bibnamefont{Naylor}},
  \bibinfo{author}{\bibfnamefont{B.}~\bibnamefont{Lee}},
  \bibinfo{author}{\bibfnamefont{B.}~\bibnamefont{Zheng}},
  \bibinfo{author}{\bibfnamefont{G.}~\bibnamefont{Liu}},
  \bibinfo{author}{\bibfnamefont{A.~T.~C.} \bibnamefont{Johnson}},
  \bibinfo{author}{\bibfnamefont{A.}~\bibnamefont{Pan}}, \bibnamefont{and}
  \bibinfo{author}{\bibfnamefont{R.}~\bibnamefont{Agarwal}},
  \bibinfo{journal}{ACS Photonics} \textbf{\bibinfo{volume}{5}},
  \bibinfo{pages}{192–204} (\bibinfo{year}{2018}).

\bibitem[{\citenamefont{Wang et~al.}(2018)\citenamefont{Wang, Chernikov,
  Glazov, Heinz, Marie, Amand, and Urbaszek}}]{Wang2018}
\bibinfo{author}{\bibfnamefont{G.}~\bibnamefont{Wang}},
  \bibinfo{author}{\bibfnamefont{A.}~\bibnamefont{Chernikov}},
  \bibinfo{author}{\bibfnamefont{M.~M.} \bibnamefont{Glazov}},
  \bibinfo{author}{\bibfnamefont{T.~F.} \bibnamefont{Heinz}},
  \bibinfo{author}{\bibfnamefont{X.}~\bibnamefont{Marie}},
  \bibinfo{author}{\bibfnamefont{T.}~\bibnamefont{Amand}}, \bibnamefont{and}
  \bibinfo{author}{\bibfnamefont{B.}~\bibnamefont{Urbaszek}},
  \bibinfo{journal}{Reviews of Modern Physics} \textbf{\bibinfo{volume}{90}},
  \bibinfo{pages}{021001} (\bibinfo{year}{2018}).

\bibitem[{\citenamefont{Kavokin et~al.}(2008)\citenamefont{Kavokin, Baumberg,
  Malpuech, and Laussy}}]{Kavokin_MCs}
\bibinfo{author}{\bibfnamefont{A.~V.} \bibnamefont{Kavokin}},
  \bibinfo{author}{\bibfnamefont{J.~J.} \bibnamefont{Baumberg}},
  \bibinfo{author}{\bibfnamefont{G.}~\bibnamefont{Malpuech}}, \bibnamefont{and}
  \bibinfo{author}{\bibfnamefont{F.~P.} \bibnamefont{Laussy}},
  \emph{\bibinfo{title}{Microcavities}} (\bibinfo{publisher}{Oxford University
  Press, Oxford, UK}, \bibinfo{year}{2008}).

\end{thebibliography}

\newpage
\section*{Appendix}

\subsection*{A. NP-free matrix}

The medium with excitons but without the NP is polarized by an external EM field that is in resonance with the excitonic transition. The polarization, $\vec {P}(\vec{r})$ is determined by Eq. (\ref{HN_MV_eq2}) with the dipole-dipole interaction tensor given by Eq. (\ref{HN_MV_eq5}).
In the limit $q \rightarrow 0$, which will be of interest here, Eq. (\ref{HN_MV_eq5}) reduces to $\vec{\nabla} \otimes \vec{\nabla}(1/{R})$. However, we shall keep it finite so far, in order to assure that the field is transverse, $\vec{{\cal E}_0}\bot \vec{q}$. 

 The polarization is a regular function of coordinates , so we apply the Fourier ttransform to it,
\begin{equation}\label{HN_MV_eq7}
\vec {P}(\vec{r}) =\int  {\vec {p}(\vec{k})e^{i\vec{k}\cdot \vec{r}}\frac {d \vec{k}}{(2\pi)^3}}\, .
\end{equation}
Substituting (\ref{HN_MV_eq7}) into Eq. (\ref{HN_MV_eq2}) yields:
$$
\int  {\vec {p}(\vec{k})e^{i\vec{k}\cdot \vec{r}}\frac {d \vec{k}}{(2\pi)^3}}=
\chi_e \left \{\int  {\vec {E_0}e^{i\vec{k}\cdot \vec{r}}\delta (\vec{k}-\vec{q})\frac {d \vec{k}}{(2\pi)^3}}+\int  {\hat {t}(\vec {k},\omega)\vec {p}(\vec{k})e^{i\vec{k}\cdot \vec{r}}\frac {d \vec{k}}{(2\pi)^3}}
\right \}\, ,
$$
where
\begin{equation}\label{HN_MV_eq9}
\hat {t}(\vec {k},\omega)=\int _{r>b} {\hat {T}(\vec {r},\omega)e^{i\vec{k}\cdot \vec{r}}{d \vec{r}}}\, .
\end{equation}
Therefore we have:
\begin{equation}\label{HN_MV_eq10}
\vec {p}(\vec{k})=\chi_e \left \{(2\pi)^3 \vec {E_0}\delta (\vec{k}-\vec{q}) + \hat {t}(\vec {k},\omega)\vec {p}(\vec{k}) \right \}\, .
\end{equation}
From Eq. (\ref{HN_MV_eq9})  we can write: 
\begin{equation}\label{HN_MV_eq11}
\vec {p}(\vec{k})\hat {t}(\vec {k},\omega)=\int _{r>b} {e^{i\vec{k}\cdot \vec{r}}
\left \{\frac{q^2e^{iqr}}{R}\vec {p}(\vec{k})+\left (\vec {p}(\vec{k})\cdot \vec \nabla \right)\vec \nabla \left ( \frac{e^{iqr}}{R} \right )
\right \}{d \vec{r}}}\, .
\end{equation}
Since the tensor $\hat {t}(\vec {k},\omega)$ is symmetric,  (\ref{HN_MV_eq9}) is just the transposed of the second term in brackets in  Eq. (\ref{HN_MV_eq10}). 

In the limit $q\rightarrow 0$, the first term in the integral in  Eq. (\ref{HN_MV_eq11}) vanishes, while ine second one we may write as:
$$
\left (\vec {p}(\vec{k})\cdot \vec \nabla \right)\vec \nabla \left ( \frac{e^{iqr}}{R} \right )
\approx e^{iqr}\left (\vec {p}(\vec{k})\cdot \vec \nabla \right)\vec \nabla \left ( \frac{1}{R} \right )\, .
$$
Therefore we have the following integral (over infinite volume):
\begin{equation}\label{HN_MV_eq13}
\vec {{\cal I}}=\int  {\left [\Theta (r-b)
e^{i(\vec{k}\cdot \vec{r}+qr)}\left (\vec {p}(\vec{k})\cdot \vec \nabla \right)\vec \nabla \left ( \frac{1}{R} \right )
\right ]{d \vec{r}}}\, ,
\end{equation}
where $\Theta $ is the Heaviside function. Integrating  (\ref{HN_MV_eq13}) by parts and using spherical trigonometry relations, we obtain (in the $q\rightarrow 0$ limit):
\begin{equation}\label{HN_MV_eq22}
\hat {t}(\vec {k},\omega)\vec {p}(\vec{k}) =\vec {{\cal I}}=\frac {4\pi}3\vec {p}(\vec{k}) 
- {4\pi}\left (\vec {p}(\vec{k}) \cdot  \vec {e_k}\right )\vec {e_k}\;,
\end{equation}  
where $\vec {e_k}={\vec {k}}/{k}$. 
For a transverse field, $ \vec {E_0} \bot \vec q$, we have
\begin{equation}\label{HN_MV_eq25}
\vec {p}(\vec{k}) =(2\pi)^3\frac{ \chi_e}{1-\frac{4 \pi}{3} \chi_e}\vec{{E} _0}\delta (\vec{k}-\vec{q})
\end{equation}  
and, by applying the inverse Fourier transform, we obtain the polarization field, $\vec {P}(\vec{r})$, given by Eq. (\ref{EP}).

\subsection*{B. Evaluation of NP's self-induced dipole moment }
 In order to evaluate the integral
$$
\vec {\cal J} = \int_{|\vec{r}|>a} \hat{T}(\vec{r})\left (\hat{T}(\vec{r})\vec{p_0} \right ) d\vec{r}\;,
$$
which represents the  self-induced dipole moment of the NP in Eq. (\ref{HN_MV_eq12}),
we denote:
$$
\vec{A}= \hat{T}(\vec{r},0)\vec{p_0}
=\frac{3(\vec{e_r}\cdot \vec{p_0})\vec{e_r}-\vec{p_0}}{r^3}\;,$$
and
$$
 \hat{T}(\vec{r})\vec{A}=\frac{3(\vec{e_r}\cdot \vec{A})\vec{e_r}-\vec{A}}{r^3}=\frac{3(\vec{e_r}\cdot \vec{p_0})\vec{e_r}+\vec{p_0}}{r^6}\;.
$$
Then we write:
$$
\int \frac{3(\vec{e_r}\cdot \vec {p_0})\vec {e_r}+\vec {p_0}}{r^6}d\vec{r}=B\vec{p_0}\; ,
$$
where
$$
B=\frac{1}{p_0^2}\int_{r>a}  \frac{3(\vec{e_r}\cdot \vec {p_0})\vec{e_r}+\vec{p_0}}{r^6}d\vec{r}
=2\pi \int_{a}^{+\infty} \int_{-1}^1 (3 \cos^2(\theta)+1) d \cos(\theta) \frac{dr}{r^4}=\frac{8 \pi}{3a^3}\;.
$$
Thus, we have:
\begin{equation}\label{A2_1}
\vec {\cal J} =\frac{8 \pi}{3a^3}\vec{p_0}\;.
\end{equation}

\subsection*{C. Parametrization of the semiconductor dielectric function}

The dielectric function of a direct band gap semiconductor, taking into account the exciton effect, is composed of two contributions, one due to unbound continuum states and other originating from discrete (bound) exciton states. 
For the former, we can write for the imaginary part:\cite{Basu} 
\begin{equation}
\displaystyle
\epsilon _{IB} ^{\prime \prime}(\omega )=\frac{4}{3} \sqrt {\text {Ry} } \left(\frac{\mu}{m_0}\right)^{3/2}E_{P} \frac{\sqrt{\hbar \omega-E_g}}{(\hbar \omega)^2}C(\omega) \,;\qquad  \hbar \omega \geq E_g\, .
\label{HN_MV_eq38}
\end{equation}
Here $m_0$ is the free electron mass,  $\mu$ is the reduced mass of the exciton, $\text {Ry}=\frac{m_0 e^4}{2 \hbar^2}=13.6\; eV$, $E_P=\frac{2 P_{cv}^2}{m_0}$ ($P_{cv}$ is the momentum matrix element between the conduction and valence bands), and  
\begin{equation}
\displaystyle
C(\omega) =\frac {2\pi \sqrt {R_x/(\hbar \omega-E_g)}}{1-\exp \left ( {-2\pi \sqrt {R_x/(\hbar \omega-E_g)}}\right )}
\,;\qquad  \hbar \omega \geq E_g\, ,
\label{HN_MV_eq41}
\end{equation}
takes into account the electron-hole interaction effect. In Eq. (\ref {HN_MV_eq41}) $R_x$ is the exciton binding energy.

The contribution due to bound excitons is given by the Elliott formula\cite{Basu} where we take into account only the ground state ($n=1)$:
\begin{equation}
\displaystyle
\epsilon _{ex} ^{\prime \prime}(\omega )=\left(\frac{2\pi e }{m_0\omega}\right )^2 \frac{P_{cv}^2}{3\pi ^2 a_x^3} \text {Im}\left(\frac 1 {\omega _0 - \omega -i\gamma } \right )\, ,
\label{HN_MV_eq42}
\end{equation}
where $a_x$ is the exciton Bohr radius, $\omega _0 =(E_g-R_x)/\hbar$ and $\gamma $ is a damping parameter. Equation (\ref {HN_MV_eq42}) may also be written in the form: 
\begin{equation}
\displaystyle
\epsilon _{ex} ^{\prime \prime}(\omega )=\eta ^2\text {Im}\left(\frac {\omega _{LT}} {\omega _0 - \omega -i\gamma } \right )\, ,
\label{HN_MV_eq43}
\end{equation}
where $\eta$ is the refractive index and $\omega _{LT}$ is called longitudinal-transverse splitting of the exciton resonance.\cite{Kavokin_MCs} It is convenient to rewrite the interband contribution also in terms of this parameter,  
\begin{equation}
\displaystyle
\epsilon _{IB} ^{\prime \prime}(\omega )=\eta ^2
\frac{\hbar \omega _{LT}\sqrt {\hbar \omega-E_g}}{2R_x^{3/2}} C(\omega) \,;\qquad  \hbar \omega \geq E_g\, .
\label{HN_MV_eq47a}
\end{equation}
Combining Eqs. (\ref {HN_MV_eq43}) and (\ref {HN_MV_eq47a}) and neglecting the contribution of the interband transitions to the real part of the dielectric function, we can write it in the following compact form: 
\begin{equation}
\displaystyle
\hat \epsilon (\omega )=\eta ^2 \left [1+\frac {\omega _{LT}} {\omega _0 - \omega -i\gamma }
+i\frac{\hbar \omega _{LT}\sqrt {\hbar \omega-E_g}}{2R_x^{3/2}} C(\omega)\Theta (\hbar \omega-E_g) 
\right ] \, .
\label{HN_MV_eq48}
\end{equation}
where $\Theta (E)$ denotes the Heaviside function. The parameters entering  Eq. (\ref {HN_MV_eq48})  are well known for the most studied semiconductors.. For instance, for GaAs we have:\cite{Basu,Kavokin_MCs} $\eta =3.3$, $R_x=4.2\, meV$, $\hbar \omega _{LT}=0.08\, meV$,  $E_g=1.51\, eV$ (at 4 K temperature).

If the excitonic effect vanishes, the second term in the brackets in Eq. (\ref {HN_MV_eq48}) disappears and $C(\omega) \rightarrow 1$ in the third one. Then it is more convenient to use the form of Eq. (\ref {HN_MV_eq38}). 
The parameters are not so well known in the case of CuO, so we take: $\eta \approx 2.5$, $E_g=1.5 \, eV$\cite{CuO} and typical values $ \mu \sim (0.1-0.2){m_0}$ and $E_P \approx 20\, eV$.

\end{document}